\documentclass[usegraphicx]{mn2e}
\usepackage{subfigure}
\usepackage{times}

\newcommand{\kms}{$\mathrm{km \, s^{-1}}$}

\title[The spin period - eccentricity relation]
      {The spin period -- eccentricity relation of double neutron stars: evidence for weak supernova kicks?}

\author[Dewi, Podsiadlowski \& Pols]
       {J. D. M. Dewi,$^{1,2}$\thanks{email: jasinta@ast.cam.ac.uk (JDMD)}, 
        Ph. Podsiadlowski,$^{2}$ O. R. Pols$^{3}$\\
        $^{1}$Institute of Astronomy, University of Cambridge,
              Madingley Road, Cambridge CB3 0HA, UK\\
        $^{2}$Department of Astrophysics, University of Oxford,
              Keble Road, Oxford OX1 3RH, UK\\
        $^{3}$Astronomical Institute, University Utrecht,
              Postbus 80000, 3508 TA Utrecht, The Netherlands}
\date{Accepted . Received ; in original form }

\pagerange{\pageref{firstpage}--\pageref{lastpage}}
\pubyear{}

\begin{document} 

\maketitle 

\label{firstpage}

\begin{abstract}
Double neutron stars (DNSs), binary systems consisting of a radio pulsar and a generally undetected second neutron star (NS), have proven to be excellent laboratories for testing the theory of general relativity. The seven systems discovered in our Galaxy exhibit a remarkably well-defined relation between the pulsar spin period and the orbital eccentricity. Here we show, using a simple model where the pulsar is spun up by mass transfer from a helium-star companion, that this relation can only be produced if the second neutron star received a kick that is substantially smaller (with a velocity dispersion of less than 50~\kms) than the standard kick received by a single radio pulsar. This demonstrates that the kick mechanism depends on the evolutionary history of the NS progenitor and that the orbital parameters of DNSs are completely determined by the evolution in the preceding helium star -- neutron star phase. This has important implications for estimating the rates of NS-NS mergers, one of the major potential astrophysical sources for the direct detection of gravitational waves, and for short-period gamma-ray bursts.
\end{abstract}


\section{Introduction}
\label{e-pspin:sec:intro}

A double neutron star (DNS) system is defined as a binary system consisting of two neutron stars (NSs) orbiting each other. One of the NSs is observed as a radio pulsar with a typical spin period of tens of milliseconds, while the other NS is usually undetectable, except in the case of the double pulsar J0737--3039 where the second NS is also observed as a slowly spinning radio pulsar (Lyne et al.\ 2004). The first DNS discovered is B1913+16 (Hulse \& Taylor 1975), and to date seven such systems are known in our Galaxy. Another DNS, B2127+11C discovered in the globular cluster M15, is excluded from this discussion as it may have a different formation mechanism. Since the original discovery, DNSs have served as a major test bed for the theory of general relativity. Measurements of the orbital decay in B1913+16 have provided the first indirect evidence for the existence of gravitational wave radiation. Gravitational waves may be detected directly in the future when two spiralling-in NSs collide with each other and merge, a phenomenon detectable by gravitational wave observatories such as LIGO, VIRGO, and LISA. The merger of two NSs is also considered a prime candidate as a source of short-duration gamma-ray bursts (Janka et al.\ 1999).

	\begin{figure}
  	 \centerline{\includegraphics[width=60mm]{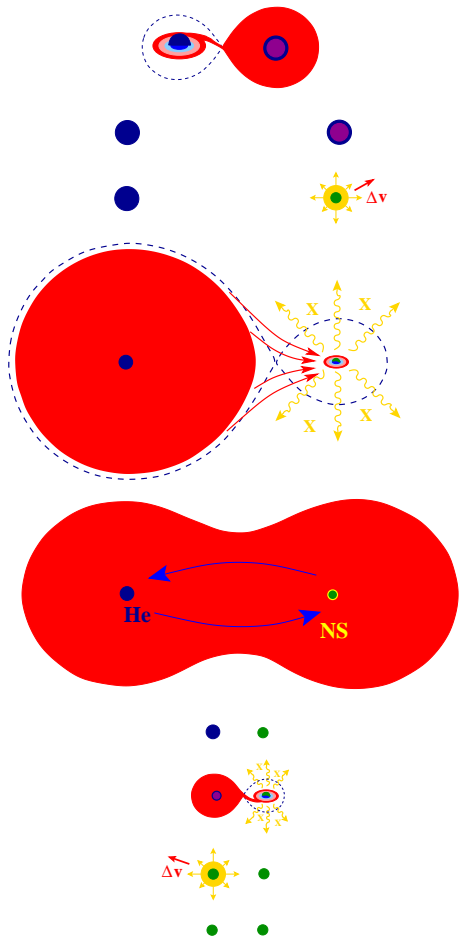}}
\caption{The formation of a double neutron star binary in the standard scenario. After transferring matter to the secondary, the primary collapses to form a neutron star, and the system evolves into a high-mass X-ray binary. The system then experiences a common-envelope phase where the neutron star spirals in inside the companion's envelope, ultimately ejecting it and leaving a short-period helium star -- neutron star binary. Mass transfer from the helium star to the neutron star spins up the neutron star, making it a recycled pulsar.  The helium star eventually explodes to become the second neutron star.}
\label{e-pspin:fig:standard}
	\end{figure}

	\begin{figure}
  	 \centerline{\includegraphics[width=60mm]{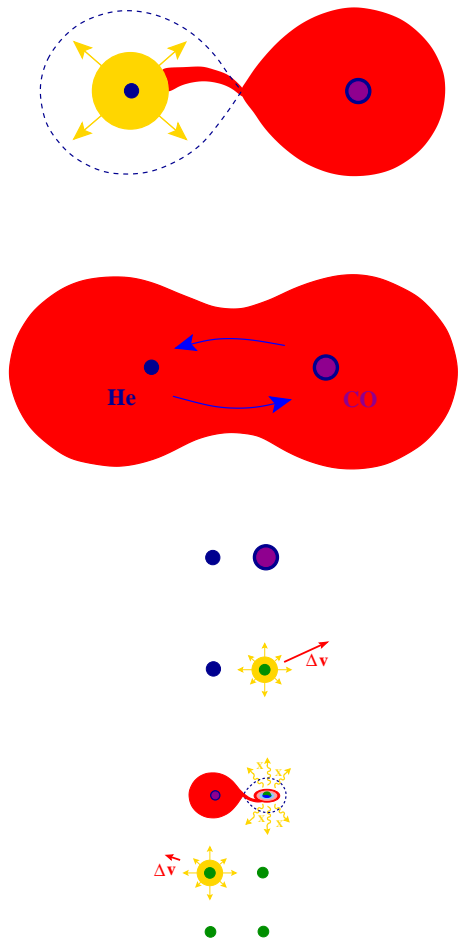}}
\caption{The formation of double neutron star binary in
the double core scenario. The initial binary consists of two
components of almost equal mass (within $\sim$ 5 -- 10 per cent). The system experiences a common-envelope phase after both stars have developed compact cores. After the ejection of the common envelope, the system has become a short-period binary consisting of two helium stars, where one has already developed a CO core (CO in the figure). The core of the primary explodes first to form the first neutron star. The subsequent evolution is almost identical to the standard scenario. Mass transfer from the helium star spins up the first neutron star, and the helium star eventually explodes to become the second neutron star.}
\label{e-pspin:fig:double}
	\end{figure}

Several theories have been proposed to explain the formation of DNSs. The most commonly discussed evolutionary channel, which we shall refer to as the standard scenario (Bhattacharya \& van den Heuvel 1991) shown in Fig.~\ref{e-pspin:fig:standard}, involves a high-mass X-ray binary (HMXB) phase at an intermediate stage. One important alternative scenario, originally proposed by Brown (1995), is the double core scenario. In this evolutionary channel, shown in Fig.~\ref{e-pspin:fig:double}, the binary initially consists of two components of almost equal masses, and a HMXB phase is avoided. Although the initial conditions and the evolutionary details in these two channels are different, the later evolutionary stages before the formation of the second NS are essentially the same, involving a helium star -- neutron star (HeS-NS) binary. Depending on the mass of the helium star and the orbital period, the helium star may transfer matter to the NS. It is this HeS-NS phase that determines the key characteristics of DNSs, in particular the spin period of the fast pulsar (22 -- 104~ms) and its relatively low magnetic fields of order $10^{10}$ G. This is believed to be the result of a recycling process, caused by the accretion of matter and angular momentum on to the first-born NS (Smarr \& Blanford 1976; van den Heuvel \& Taam 1984), spinning up the NS and possibly driving the decay of the NS's magnetic field (by a not-yet-understood mechanism; Srinivasan \& van den Heuvel 1982). In contrast to these recycled pulsars, typical isolated pulsars have long spin periods of order 1~s and magnetic field strengths of order $10^{12}$ G.

Observations of young {\em single} radio pulsars (Lyne \& Lorimer 1994) have shown that most NSs receive a large kick velocity when they are born with a mean birth velocity of $\sim$ 200 -- 450~\kms\ (e.g., Hansen \& Phinney 1997; Cordes \& Chernoff 1998; Arzoumanian, Chernoff \& Cordes 2002), presumably due to an asymmetry in the supernova (SN) explosion (e.g., Shklovskii 1970).  The study by Arzoumanian et al.\ (2002) found a bimodal Gaussian distribution with velocity dispersions of 90~\kms\ and 500~\kms\ for the two components, respectively, where about 50 per cent of pulsars have velocities greater than 500~\kms\ and only 10 per cent have velocities less than 100~\kms. The most recent study by Hobbs et al.\ (2005), which was based on a much larger sample of proper motion measurements of pulsar, revealed a distribution with dispersion of 265~\kms, with no evidence for a bimodal distribution.

If the NS is in a binary system, the kick velocity affects the orbital parameters of the post-SN system, in particular its eccentricity and the system space velocity as a function of orbital period (Hills 1983; Brandt \& Podsiadlowski 1995).  Large kicks tend to produce a wider range of eccentricities and orbital periods and larger system velocities. In contrast, in a symmetric SN, the post-SN orbital parameters are entirely determined by the amount of mass lost during the explosion, where more mass loss leads to wider orbits and larger eccentricities. Most importantly, the observed distributions of post-SN binary parameters  can be used to constrain the SN kick distribution (e.g., Brandt \& Podsiadlowski 1995; Pfahl et al.\ 2002). Several studies have been carried out to constrain the kick velocity required in the formation of the observed DNSs; either solely from their orbital parameters (e.g., Fryer \& Kalogera 1997; Dewi \& Pols 2003; Dewi \& van den Heuvel 2004; Willems \& Kalogera 2004), or from additional information such as the proper motion of the pulsars and the inferred misalignment angles between the orbit and spin axes (e.g., Wex, Kalogera \& Kramer 2000; Willems, Kalogera \& Henninger 2004; Thorsett, Dewey \& Stairs 2005). In this paper we propose a new constraint on the SN kick velocity, which is based solely on the observed correlation of pulsar spin period and the orbital eccentricity of the systems.

	\begin{figure*}
         \mbox{\subfigure[$\sigma_2 =$ 190~\kms]{\includegraphics[width=60mm]{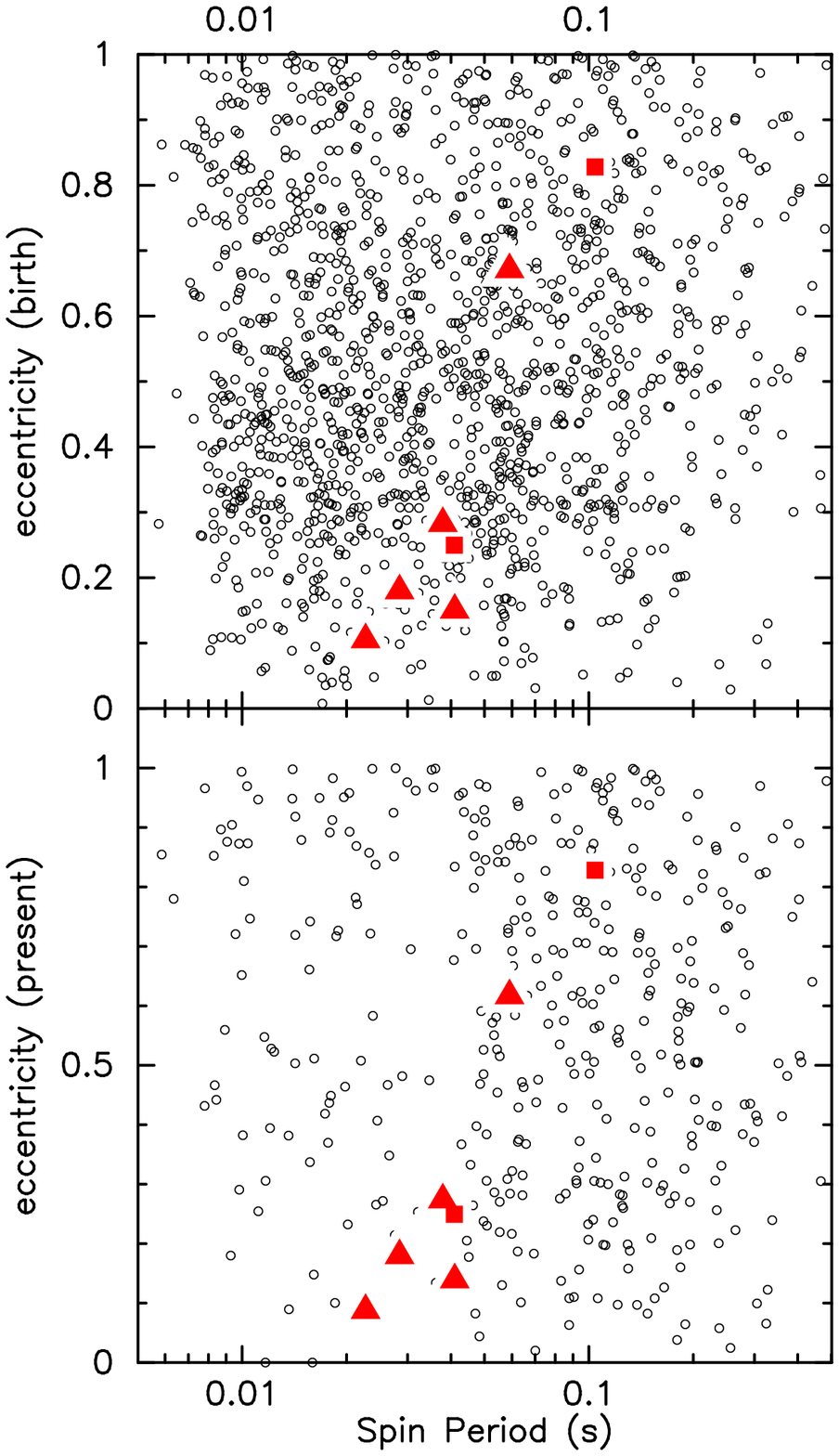}}
               \quad \quad
               \subfigure[$\sigma_2 =$  20~\kms]{\includegraphics[width=60mm]{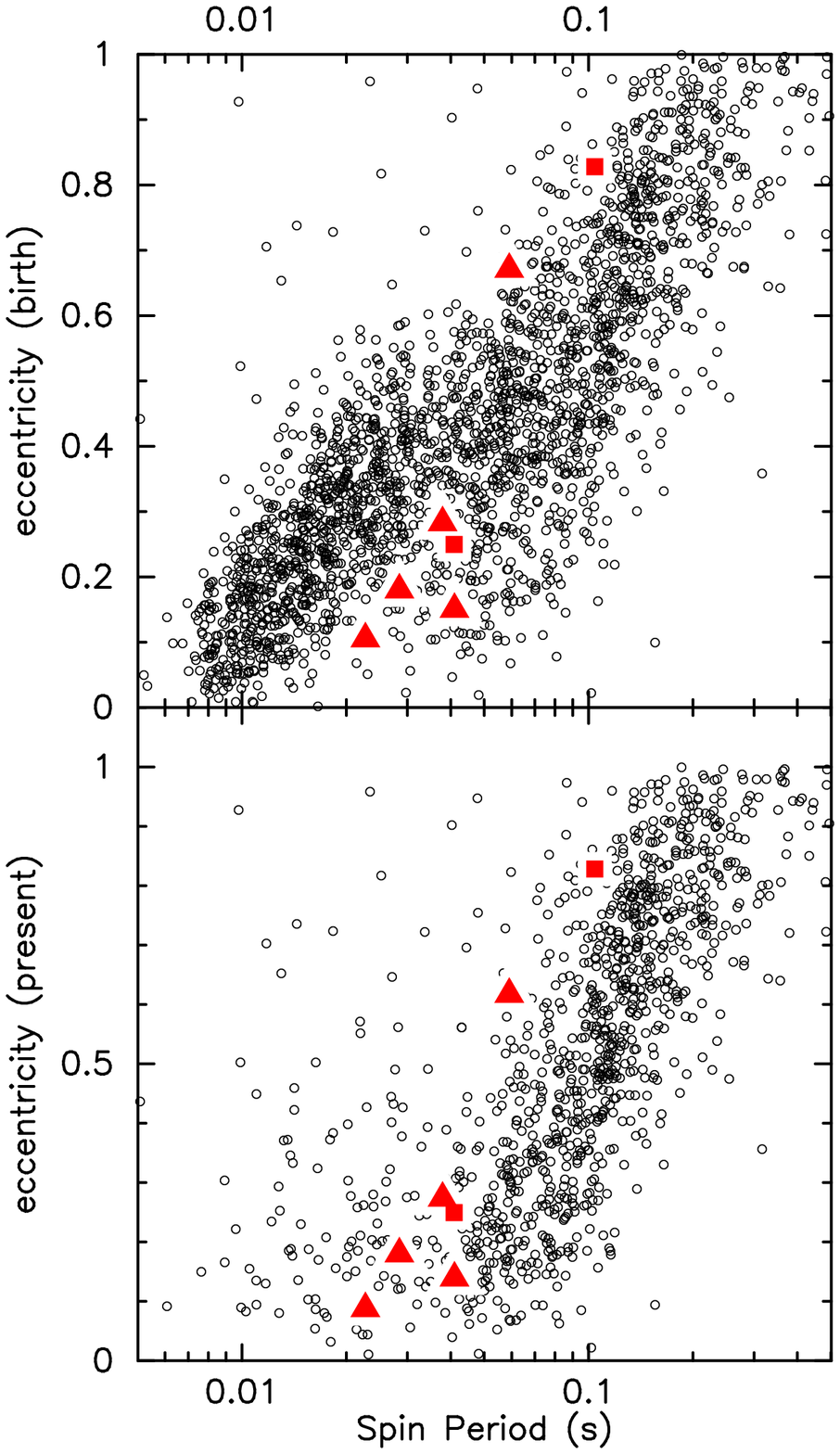}}}
\caption{Population synthesis results showing a scatter plot (open circles) of the theoretical relation between the spin period vs.\ eccentricity in the double core scenario for two kick distributions for the second-born neutron star: a conventional kick distribution with a velocity dispersion of 190~\kms\ (left panels) and a low-velocity distribution with a velocity dispersion of 20~\kms\ (right panels). The plot present distributions at birth (upper panels) and at present (lower panels) after taking into account the orbital decay due to gravitational-wave radiation. Red symbols represent the observed properties of seven Galactic double neutron stars: squares indicate possible wind-accreting systems; triangles show mass-accreting systems.}
\label{e-pspin:fig:spin}
	\end{figure*}

\section{Spin period versus eccentricity relation}
\label{e-pspin:sec:relation}

After seven galactic DNSs have been discovered, a clear relation between the spin periods, $P_{\mathrm{s}}$, of the first-born pulsar and the orbital eccentricities, $e$, has become apparent (McLaughlin et al.\ 2005; Faulkner et al.\ 2005). Faulkner et al.\ suggested that this relation has its origin in the fact that in HeS-NS binaries where the helium star transfers a larger amount of mass to the NS, the NS is spun up to a shorter spin period, but that in turn these system experience less mass loss in the SN, leading to a lower eccentricity. However, this argument does not consider the effect of a SN kick, which in general will have the tendency to randomize any correlation of this type.

To investigate whether the existence of this $P_{\mathrm{s}}-e$ relation is physical or merely a coincidence, we performed population synthesis studies of the formation of DNSs, applying both the standard channel (Dewi, Pols \& van den Heuvel, in preparation, hereafter DPvdH) and the double core scenario (Dewi, Podsiadlowski \& Sena, in preparation, hereafter DPS). We assume that the NS accretes matter mainly from the helium star during the mass-transfer phase in HeS-NS binaries, and that the accretion rate is limited by the Eddington rate for helium accretion, $\dot{M}_{\mathrm{Edd}}$. This is a reasonable assumption as the majority of the DNS population (more than 80 per cent in the typical cases) is formed through a mass transfer phase (DPvdH). Before mass accretion, we assume that a NS has a spin period of 1 s. The process of spin up is not fully understood; here we model the spin up by making a simple, plausible assumption that the accreted matter brings with it a total angular momentum of $\Delta J = \sqrt{G M R_{\mathrm{A}}} \, \Delta M$, where $\Delta M = \dot{M}_{\mathrm{Edd}} \Delta t$ and $\Delta t$ is the duration of the mass-transfer phase.  $R_{\mathrm{A}}$ is the Alfv\'en radius, defined as the radius within which matter is coupled to the magnetic field of the NS and starts to corotate with the NS magnetosphere. $R_{\mathrm{A}}$ depends on the magnetic field strength $B$ of the NS, and we assume for simplicity that after the accretion, $B$ has decayed to $10^{10}$ G, a value that is close to the average value of the magnetic field strengths of the observed first-born NSs in DNSs. The change in spin angular momentum $\Delta J$ is then applied to derive the recycled spin period of the NS.  To simulate the distribution of spin periods and orbital parameters for a population of DNSs, we employed the results of a systemic study of the evolution of HeS-NS binaries (Dewi et al.\ 2002; Dewi \& Pols 2003).

Following the standard arguments that NSs are born with a significant kick velocity, we first applied a Maxwellian kick velocity distribution with a dispersion $\sigma =$ 190~\kms\ (Hansen \& Phinney 1997) at the birth of both NSs. The upper-left panel in Fig.~\ref{e-pspin:fig:spin} shows the resulting distribution of the eccentricity as a function of spin period for the resulting DNS systems at birth. As can be clearly seen, the effect of the SN kick has been to completely randomize the distribution in the $P_{\mathrm{s}}-e$ plane, implying that there should be no relation between the two parameters. However, this panel shows the distribution at the birth of DNSs.  Chaurasia \& Bailes (2005) suggested that the observed $P_{\mathrm{s}}-e$ correlation could be the result of a selection effect. They pointed out that DNSs in short-period highly eccentric orbits merge on a relatively short timescale due to gravitational-wave radiation and therefore are less likely to be observed at the present time. Therefore, they argue, the systems we observe are preferentially the close-orbit, low-eccentricity and wide-orbit, high-eccentricity systems, where the former are expected to have shorter spin periods because the mass-transfer phase in close binaries lasts longer. To test this hypothesis, we plot in the lower-left panel of Fig.~\ref{e-pspin:fig:spin}~ the expected current distribution of DNSs, taking into acount the orbital decay due to the emission of gravitational-wave radiation, where the distribution of ages of the DNSs since birth was assumed to be uniform between 0 and $8\times10^9$~years (the time for a spun-up pulsar with $B = 10^{10}$~G to reach the pulsar death line). We would like to point out that our procedure for taking into account gravitational-wave radiation evolution does indeed result in a depletion of systems that have short merger times: by assigning an age from a flat age distribution, all those systems with ages exceeding their merger timescale will be depleted from the population. However, while the lower-left panel indeed shows that many of the systems with short spin periods are removed (because they tend to have shorter merger times), the eccentricity distribution at a given spin period is still essentially random. Thus, correcting for orbital decay due to gravitational-wave radiation does not remove all DNSs with high eccentricity and short spin period. The reason is that these systems tend to have long orbital periods and hence the eccentricity does not significantly decay. This demonstrates that the selection effect proposed by Chaurasia \& Bailes (2005) cannot be the main explanation for the observed correlation.

It has already been suggested (Tauris \& Bailes 1996; Pfahl et al.\ 2002; Podsiadlowski et al.\ 2004) that not all NSs may receive a standard high kick velocity, in particular if they are formed in close binaries. To consider this possibility, we next assume that the second NSs (hereafter NS2) were born with low kick velocities, applying a velocity dispersion $\sigma_1 =$ 190~\kms\ for the first NS (hereafter NS1) and $\sigma_2 =$ 20~\kms\ for NS2. With this combination of kick velocity distributions, we are able to reproduce the $P_{\mathrm{s}}-e$ relation, as is shown in the right panels in Fig.~\ref{e-pspin:fig:spin}. This is quite remarkable considering the simplicity of our spin-up model: the slope of the correlation is reproduced extremely well in the expected current distribution (lower right panel). The case of a low kick velocity is similar to the case of a symmetric SN explosion, where the orbital parameters after the SN are entirely determined by the mass loss during the explosion. Hence, lower SN mass loss results in a lower eccentricity and a shorter orbital period.  The amount of mass lost in the explosion depends entirely on the previous evolution of the HeS-NS binaries. The systems with lower mass loss in the SN are those that had lower initial HeS masses; but these systems also had a longer timescale for the mass-transfer phase. Since the amount of spin-up depends essentially on the duration of the mass-transfer phase, a longer mass-transfer phase leads to a shorter spin period for the recycled pulsar, explaining the observed relation. With a low $\sigma_2$, we do not produce systems with short orbital period and high eccentricity (DPvdH; DPS); all systems with high eccentricity are in wide orbit. Hence, the population of high-eccentricity systems in the lower-right panel of Fig.~\ref{e-pspin:fig:spin} is essentially the same as in the upper-right panel.

To also investigate the effect of the kick magnitude at the birth of NS1, we applied different kick velocity distributions for the first SN, using $\sigma_1 =$ of 90 and 500~\kms\ (Arzoumanian et al.\ 2002), respectively, while keeping $\sigma_2 =$ 20~\kms. We found that the $P_{\mathrm{s}}-e$ relation is retained, irrespective of the magnitude of the kick velocity of the first SN. This implies that the orbital parameters are determined only by the second SN. Next, to study the importance of the kick at the birth of NS2, we varied $\sigma_2$ from 0 to 100~\kms, and found that the $P_{\mathrm{s}}-e$ relation is maintained provided that $\sigma_2$ is less than 50~\kms, which we consider an upper limit on a Maxwellian velocity dispersion for the second SN kick, substantially lower than the standard kick.

The result shown in Fig.~\ref{e-pspin:fig:spin} is for a simulation of the double core scenario. Although some of the details of evolution in the standard channel are different, the latter scenario yields the same result as the double core scenario, i.e.\ the $P_{\mathrm{s}}-e$ relation exists as long as $\sigma_2$ is low, irrespective of $\sigma_1$. As the two channels share the same final stages of evolution, we can conclude that the evolution of HeS-NS binaries is the most important stage which will determine the physical properties of DNSs.

We note, however, that the work discussed so far was based on the assumption that the NS only accretes matter during a mass-transfer phase in HeS-NS binaries. On the other hand, DPvdH find that two out of 7 galactic DNSs (i.e.\ J1811--1736 and J1518+4904) have not experienced a mass-transfer phase in the HeS-NS stage (these systems have been indicated as squares in Fig.~\ref{e-pspin:fig:spin}). Even though these systems may not experience standard mass accretion, the first-born NS may still accrete matter from the wind of the helium star, or from the stellar wind of and the transferred matter from the giant secondary (in the case of the standard scenario). As a detailed study for this wind accretion case is not yet available, we will leave the discussion for further study. However, we argue that these systems may also produce a similar $P_{\mathrm{s}}-e$ relation, but contribute only to the upper-right part of the relation. The reason for this is that wind mass loss tends to decrease the stellar mass, but not as much as mass transfer, and hence tends to widen the orbit; therefore at the moment of the collapse of the second star, the star still has a relatively large mass and is in a wide orbit. Hence, we expect the resulting DNSs to have long orbital periods and large eccentricities. On the other hand, the NS does not accrete much matter from the wind and therefore is only mildly recycled with spin periods not much less than the original.

Based on the space velocities of three DNSs, the necessity of low kick velocities for some systems has been proposed before (Hughes \& Bailes 1999). In order to reconcile this result with the significant natal kick hypothesis, Hughes \& Bailes (1999) sceptically suggested that errors in the pulsar velocity measurements may give rise to the inferred low kick velocity. Another indication that a low kick velocity is required for the second NSs comes from the fact that five out of seven DNSs have eccentricities lower than 0.3 (van den Heuvel 2004). Chaurasia \& Bailes (2005) attributed this to a selection effect against discovering short-period highly eccentric systems because they merge on a relatively short timescale or evolve to low eccentricities. However, our results show that this selection effect alone is not sufficient to reproduce the observed distribution, and that low kick velocities are indeed required. Systems with low eccentricities originate from HeS-NS binaries with low mass helium stars and short orbits. Due to the selection effect on the initial mass function and initial period distribution, we expect to find more of these systems, and hence more low-eccentricity DNSs are produced. However, we realize that whether the bias against discovering higher-eccentricity is merely due to low kick velocity still needs further investigation, which is beyond the aim of this paper. The need of low kick velocity is in sharp contrast to the general current belief that NSs are born with a significant kick velocity. If the observed relation between spin period and eccentricity holds up, it is the first firm evidence that clearly rules out a standard kick for the second NS in DNSs, fundamentally changing our understanding of the NS kick origin and mechanism. One possible explanation for such a low kick velocity is a low pre-SN core mass of the helium star before the explosion (Tauris \& Bailes 1996; Podsiadlowski et al.\ 2004), as may preferentially occur in close binaries and as is directly applicable to the short-period HeS-NS binaries relevant for DNSs. 

For two DNSs, misalignment angles between the spin and orbital angular momentum vectors have been derived using pulse-profile-related methods, i.e.\ $22^{\circ} \,$ for B1913+16 (Kramer 1998; Weisberg \& Taylor 2002) and $25^{\circ} \,$ for B1534+12 (Stairs, Thorsett \& Arzoumanian 2004). The asymmetric first SN explosion will cause the spin angular momentum of NS1 to be misaligned with the orbital angular momentum. Mass transfer on to NS1 is expected to remove the misalignment. Hence, in a simple model for the pulsar spin-up we would expect the spin axis to be aligned with the orbital axis before the second SN and that this alignment would be affected by the second SN. In view of this assumption, the derived misalignment angles of B1913+16 and B1534+12 have been considered to be the signature for a significant kick velocity at the birth of the second NS. Based on the above-mentioned misalignment angles, Willems et al.\ (2004) found that a minimum kick of 190~\kms\ is required for B1913+16, while Thorsett et al.\ (2005) obtained 170~\kms\ for B1534+12. However, at present there is no evidence that the spin and orbital angular momentum vectors before the second SN must be aligned, and indeed the accretion on to magnetic neutrons stars with mis-aligned spin axes has not been properly modelled to date. Note also that the derivation of misalignment angles are model dependent. Therefore, whether or not the observed misalignmnents are entirely due to a supernova kick is still an open question, and hence our result does not necessarily contradict the previous studies by, e.g., Willems et al.\ (2004) or Thorsett et al.\ (2005).

\section{Conclusions}
\label{e-pspin:sec:conclusion}

We have performed a population synthesis study to examine the relation between the spin period and the orbital eccentricity observed among the Galactic DNSs, applying a simple model of spin-up due to accretion. We assume the first-formed neutron star accretes matter from its helium-star companion by mass transfer. We find that the relation can be reproduced only if the birth of the second neutron star in DNSs is accompanied only by a moderate supernova kick and that the physical parameters of DNSs are determined by the evolution of HeS-NS binaries. Although it is not yet evident what is responsible for the low kick velocity at the birth of the second neutron star, it strongly indicates that the kick mechanism depends very much on the evolutionary history of the NS progenitor. This will have important implications for estimating the birthrates of NS-NS mergers and short-duration gamma-ray bursts. With a low kick velocity, the probability for the progenitors of DNSs to be disrupted is also low. This may explain why fewer than 10 per cent of the single radio pulsars have kick velocities less than 100~\kms\ (Arzoumanian et al.\ 2002).

\section*{Acknowledgments}
JDMD acknowledges a Talent Fellowships from the Netherlands Organization for Scientific Research (NWO) for her stay at the University of Oxford where this work was initiated. It is a pleasure to thank Matthew Bailes, Norbert Langer, Thomas Maccarone, Andrew Melatos, Jim Pringle, Ingrid Stairs for fruitful discussions; and David Champion for releasing the data on orbital derivatives of J1829+2456 prior to publication.

\label{lastpage}

\end{document}